\documentclass{svproc}
\usepackage{subfigure}
\usepackage{slashed}
\usepackage{amsmath,amssymb}
\usepackage{graphicx}
\usepackage{units}
\usepackage{bbold}
\usepackage{xcolor}
\usepackage{dsfont}
\usepackage[hyperfootnotes=false,colorlinks,citecolor=blue]{hyperref}
\usepackage{comment}
\usepackage[normalem]{ulem} 
\usepackage{comment}
\usepackage{autobreak}
\usepackage{wrapfig}
\usepackage{cite}
\usepackage{url}
\usepackage{multicol}
\usepackage{framed}

\begin{document}
\mainmatter             
\title{Neutrino Theory Overview}
\titlerunning{Neutrino Theory Overview}  
\author{P. S. Bhupal Dev}
\authorrunning{P.S.B. Dev} 
\institute{Department of Physics and McDonnell Center for the Space Sciences, \\Washington University, St.~Louis, MO 63130, USA\\
\email{bdev@wustl.edu}
}

\maketitle              %
\vspace{-0.3cm}
\begin{abstract}
The neutrino sector is the least known in the Standard Model. We briefly review the open questions in neutrino physics and the future experimental prospects of answering them. We argue that a synergistic approach at multiple frontiers is needed.  

\keywords{Neutrinos, Beyond Standard Model Physics}
\end{abstract}
%
\section{Introduction}
\vspace{-0.2cm}
The Standard Model (SM) 
has made remarkably accurate predictions about fundamental particle properties and interactions, which have been experimentally confirmed with high precision. Nevertheless, there exist certain phenomena which cannot be explained within the framework of SM, and therefore, necessitate the existence of some beyond-the-SM (BSM) physics. A prime example comes from the neutrino sector. On one hand, neutrinos have been key in establishing two of the most important features of the SM, i.e., (i) parity breaking in weak interactions, and (ii) 3-fold repetition of the fermion family structure. On the other hand, the observation of neutrino oscillations implies that the neutrino flavor eigenstates are a linear superposition of their mass eigenstates, and at least two of the mass eigenvalues must be nonzero in order to explain the observed solar and atmospheric mass-squared differences. It necessarily requires BSM physics, because within the SM with $SU(3)_c\times SU(2)_L\times U(1)_Y$ gauge structure:  
\begin{itemize}
    \item Neutrinos are purely left-handed ($\nu_L$), and there is no right-handed partner ($\nu_R$) to write the Dirac mass term $m_D\bar{\nu}_L\nu_R$, unlike for the charged fermions. 
    \item A Majorana mass term $m_M\bar{\nu}_L^C\nu_L$ (where $\nu_L^C\equiv \nu_L^TC^{-1}$, with $C$ being the charge conjugation matrix) transforms as a triplet under $SU(2)_L$, and hence, breaks the $SU(2)_L$-gauge-invariance, apart from breaking the global lepton number symmetry ($L$) of the SM.  
    \item The non-perturbative electroweak sphaleron processes do break lepton number, but only together with baryon number ($B$) in the combination $B+L$, and still preserve the accidental global $B-L$ symmetry of the SM, thus forbidding neutrino mass generation. 

    \item Non-perturbative quantum gravity effects are expected to break all global symmetries, including the $B-L$ symmetry of the SM, but the neutrino masses generated from such effects will be at most of order $v^2/M_{\rm Pl}\sim 10^{-5}$ eV (where $v$ is the electroweak vacuum expectation value and $M_{\rm Pl}$ is the Planck mass), not enough to explain the observed mass-squared differences  $\Delta m^2_{\rm sol}\simeq 7.5\times 10^{-5}~{\rm eV}^2$ and $\Delta m^2_{\rm atm}\simeq 2.5\times 10^{-3}~{\rm eV}^2$. 
\end{itemize}

\vspace{-0.2cm}
\section{$\nu$ Wishlist}
\vspace{-0.2cm}
The 3-neutrino oscillation paradigm is entering a new precision era, with the ongoing and planned oscillation experiments poised to measure the oscillation parameters to an unprecedented accuracy~\cite{Denton:2022een}. This will be a huge step forward in our understanding of the neutrino sector, but we want to learn a lot more about them. Here is a partial list of open questions in neutrino physics: 
\begin{framed}
\vspace{-0.5cm}
\begin{enumerate}
\item \textcolor{green!50!black}{What is the neutrino mass ordering? Normal or inverted?} 
\item \textcolor{green!50!black}{In which octant is the atmospheric mixing angle?} 
\item \textcolor{green!50!black}{Is there leptonic CP violation? }
\item \textcolor{orange}{What is the absolute neutrino mass scale?} 
\item \textcolor{orange}{Are there other species of (sterile) neutrinos? }
\item \textcolor{orange}{How do neutrinos get mass? Is it Dirac or Majorana? } 
\item \textcolor{red}{Why neutrino mixing is so different from quark mixing?} 
\item \textcolor{red}{Do neutrinos decay? What is their lifetime? }
\item \textcolor{red}{Do neutrinos have non-standard interactions?} 
\item \textcolor{red}{Are neutrinos responsible for the observed baryon asymmetry?} 
\item \textcolor{red}{Do neutrinos have anything to do with Dark Matter (DM)?} 
\end{enumerate}
\vspace{-0.5cm} 
\end{framed}
I have divided these questions into three categories. The first set of 3 questions (in green) will most likely be answered by the ongoing (T2K, NO$\nu$A, IceCube, KM3NeT) and planned (JUNO, DUNE, Hyper-K) oscillation experiments within the next 10--15 year timescale. The next set of 3 questions (in orange) might be answered in the foreseeable future, if nature cooperates. For instance, the absolute neutrino mass scale can be directly probed by measuring the end point of the $\beta$-decay spectrum. Using tritium $\beta$-decay, KATRIN has set the best direct bound of $m_\nu\leq 0.45$ eV at 90\% CL, and future experiments like Project-8 aim to reach 0.04 eV. Similarly, cosmological observables set an indirect upper limit  on the sum of neutrino masses: $\sum_i m_i< 0.12 $ eV from Planck and $< 0.072$ eV from DESI, which already disfavor the inverted mass ordering. The remaining questions mentioned above (in red) are much harder to answer definitively, but within the context of a specific neutrino mass model, may be addressed to some extent. For instance, the difference between the quark and lepton mixing matrices can be understood by invoking some flavor symmetry. Although the models based on simple flavor ansatz, like tri-bi-maximal, are already ruled out by the oscillation data, there exist other flavor symmetries (e.g. $A_4$, $S_4$, $T'$, $\Delta(27)$) which are still compatible with the current data and can  be further tested with future precision oscillation data~\cite{Chauhan:2023faf}.

\vspace{-0.2cm}
\section{Majorana or Dirac?}
\vspace{-0.2cm}
As for the nature of neutrino mass, the distinction between Dirac and Majorana neutrinos is in general suppressed by some power of $m_\nu/E_\nu$, which makes it unobservable in most experimental settings with relativistic neutrinos. A `smoking gun' signal of Majorana neutrinos would be the observation of a lepton number violating (LNV) process, such as the 
neutrinoless double beta decay ($0\nu\beta\beta$)~\cite{Agostini:2022zub}. The current lower limit on the $0\nu\beta\beta$ lifetime is $T_{1/2}^{0\nu}\gtrsim 10^{26}$yr, corresponding to an upper limit on the effective Majorana mass $m_{\beta\beta}\lesssim 0.1$ eV. 
The next-generation ton-scale experiments (such as LEGEND and nEXO) are expected to reach a half-life sensitivity of $10^{28}$yr, corresponding to $m_{\beta\beta}\lesssim 0.02$ eV. However, there is no guarantee that a positive signal of $0\nu\beta\beta$ will be observed, especially if neutrino masses turn out to follow normal ordering, which is, in fact, preferred over the inverted ordering in current global oscillation fits, as well as from cosmology constraints. 

An alternative (albeit indirect) probe of the Majorana nature of neutrinos is the search for LNV in high-energy collider experiments~\cite{Deppisch:2015qwa}. Although it is a promising avenue, no such LNV signal has been observed so far, setting stringent limits on possible LNV operators. 

Another potential way to overcome the $m_\nu/E_\nu$ suppression for Majorana versus Dirac distinction is to use nonrelativistic neutrinos. There does  exist a natural source of nonrelativistic neutrinos (at least for two of them) in the form of the cosmic relic neutrino background; however, its direct detection is an extremely challenging problem in itself~\cite{Vogel:2015vfa}.  

\vspace{-0.2cm}
\section{Sterile Neutrinos}
\vspace{-0.2cm}
Perhaps the simplest possibility to generate neutrino mass is by adding right-handed neutrinos $\nu_R$ (also known as sterile neutrinos, heavy neutrinos, or heavy neutral leptons)  to the SM, which enables us to write the Dirac mass term $m_D\bar{\nu}_L\nu_R$. These additional neutrino species must be sterile (i.e. have no direct weak interactions) because of the LEP constraint on invisible $Z$ decays. Since the sterile neutrinos are SM gauge singlets, they are allowed to have a Majorana mass term $m_N\bar{\nu}_R^c\nu_R$ which, together with the Dirac mass term, leads to the neutrino mass matrix
 $   M_\nu = \begin{pmatrix}
        0 & m_D \\ m_D^T & m_N
    \end{pmatrix}$.
In a bottom-up phenomenological approach, the scale of $m_N$ is unknown. If $||m_N||\gg ||m_D||$, the light neutrino mass matrix becomes $m_\nu\simeq -m_D m_N^{-1}m_D^T$, famously known as the (type-I) seesaw formula, and neutrinos are Majorana particles. In the opposite limit,  $||m_N||\ll ||m_D||$, neutrinos become quasi-Dirac, which are fundamentally Majorana but behave like Dirac in most experimental settings due to the tiny active-sterile mass splitting $\delta m\propto m_N$. If $m_N=0$ is enforced by some symmetry, neutrinos are exactly Dirac, which is a boring but valid option. 

Experimentally, the most relevant quantity for sterile neutrino searches is the active-sterile mixing $U_{\ell N}$, along with the scale of $m_N$. In the naive seesaw limit, they are related: $U_{\ell N}\simeq m_Dm_N^{-1}\simeq (m_\nu m_N^{-1})^{1/2}$. However, this relation is strictly valid only in the single-generation case. With more than one generations, `large' mixing is allowed even for low-scale $m_N$, while being consistent with the seesaw formula, by choosing special textures of $m_D$ and $m_N$. Variations of the type-I seesaw with additional particles (such as inverse seesaw, linear seesaw, double seesaw) can also realize such low-scale seesaw with `large' mixing in a technically natural way.  

\begin{figure}[t!]
    \vspace{-0.4cm} \includegraphics[width=1.0\linewidth]{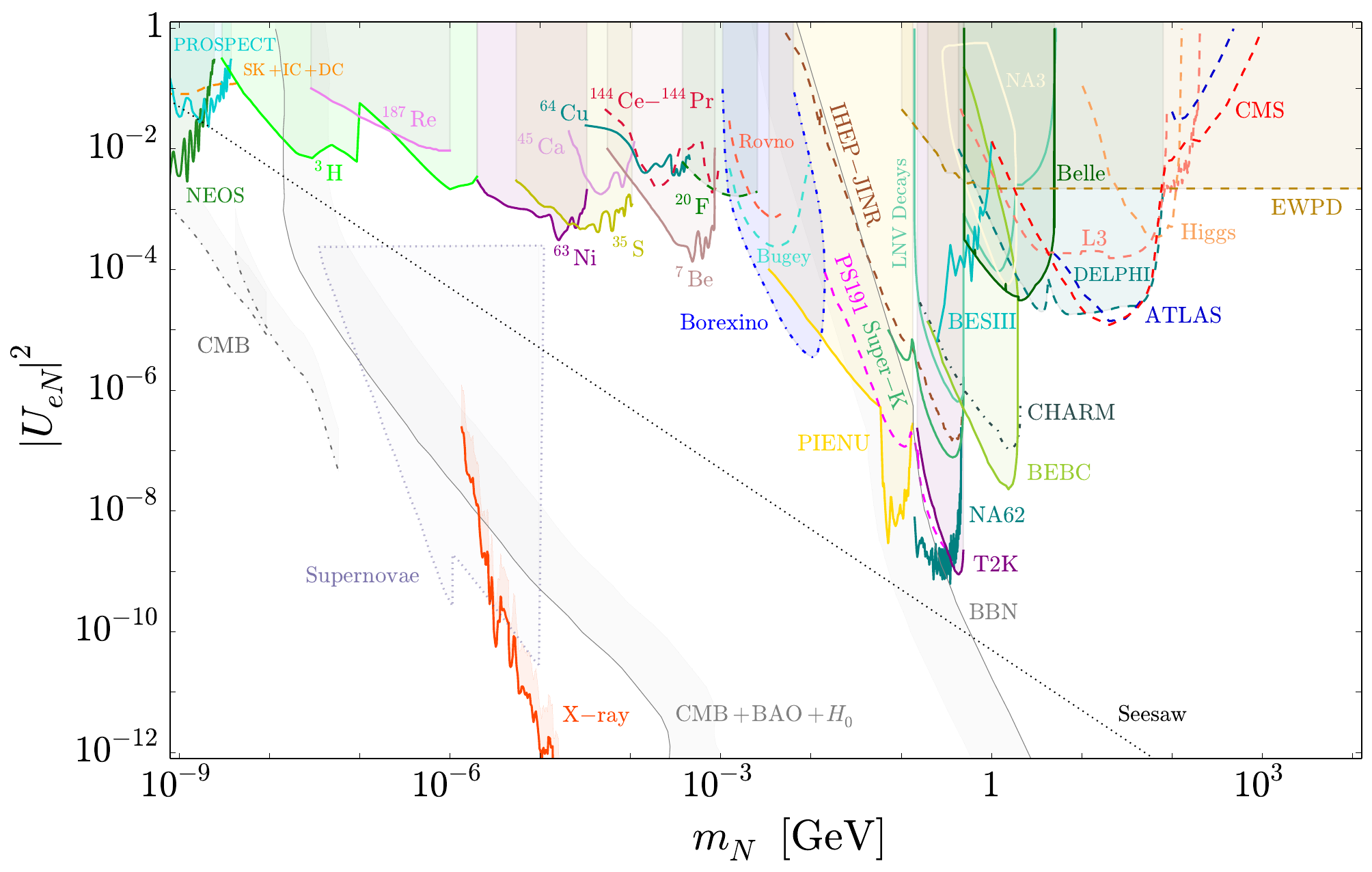}
    \vspace{-0.8cm} 
    \caption{{\it Current constraints on the active-sterile mixing strength in the electron flavor as a function of the sterile neutrino mass. Adapted from Refs.~\cite{Bolton:2019pcu}.}}
\label{fig:sterile}
\vspace{-0.8 cm} 
\end{figure}

A summary of the existing constraints on the sterile neutrino mass and mixing is shown in Fig.~\ref{fig:sterile}. Here we only consider the mixing with electron flavor; similar plots for the muon and tau flavors, as well as future experimental prospects, can be found in Ref.~\cite{Bolton:2019pcu}. The entire mass range shown here can be divided into 4 sectors, depending on which type of constraint is dominant: (i) astrophysical and cosmological constraints in the sub-MeV range, (ii) beam dump and meson decay constraints in the MeV--GeV range; (iii) collider constraints in the GeV--TeV range; and (iv) electroweak precision  constraints beyond TeV. It is worth mentioning that in the electron mixing case, although the $0\nu\beta\beta$ constraint is usually taken to be the most stringent, it can be significantly weakened depending on the sterile mass hierarchy and the relative phases between them~\cite{Bolton:2019pcu}, without 
affecting most of the other laboratory constraints. Similarly, the cosmological constraints, especially from BBN and CMB, could be weakened in presence of additional (dark) interactions of the sterile neutrino~\cite{Dasgupta:2021ies}.  Therefore, it is important to search for the sterile neutrinos in as many independent ways as possible, even if the projected sensitivities in some cases may not be as competitive as the existing constraints.   

\begin{wrapfigure}{r}{0.45\textwidth}
\centering
  \vspace{-0.5cm}    \includegraphics[width=1.0\linewidth]{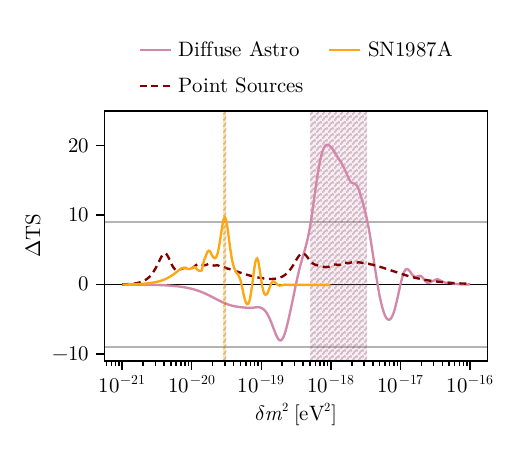}
  \vspace{-1.0cm} 
    \caption{{\it Constraints on the quasi-Dirac neutrino parameter space from astrophysical and supernova neutrino data. Adapted from Ref.~\cite{ Carloni:2025dhv}.}}
    \label{fig:QDino}
    \vspace{-1.5cm} 
\end{wrapfigure}

If neutrinos are quasi-Dirac with hyperfine mass-squared splitting $\delta m^2$ between active and sterile components, they can be probed using oscillations over astrophysical baselines. Stringent upper limits on $\delta m^2_{1,2}\lesssim 10^{-12}~{\rm eV}^2$ and $\delta m_3^2\lesssim 10^{-5}~{\rm eV}^2$ have been derived using solar, atmospheric, accelerator and reactor neutrino data. Similarly, the supernova neutrino data from SN1987A was used to rule out narrow band of $\delta m^2\in [2.5, 3.0]\times 10^{-20}~{\rm eV}^2$~\cite{Martinez-Soler:2021unz}. Recently, using the high-energy astrophysical neutrino data from IceCube, a wider range of $\delta m^2$ from $2\times 10^{-19}$ to $3\times 10^{-18}~{\rm eV}^2$ has been excluded~\cite{Carloni:2025dhv}, while there exists a preference for $\delta m^2\simeq 1.9\times 10^{-19}~{\rm eV}^2$ at $2.8\sigma$ (see Fig.~\ref{fig:QDino}), primarily driven by 
some tension between the cascade and track measurements below 30 TeV. It is interesting to note that certain string theory landscape constructions, such
as Swampland, predict that neutrinos are Dirac-like particles. Moreover, the global lepton number symmetry is expected to be broken by quantum gravity, thus turning Dirac neutrinos into quasi-Dirac.  

\vspace{-0.2cm}
\section{Going Beyond Steriles}
\vspace{-0.2cm}
There are many other ways to generate neutrino masses without involving the sterile neutrinos. At tree level, the dimension-5 Weinberg operator $LLHH/\Lambda$ can be realized by adding $SU(2)_L$-triplet scalars (type-II seesaw) or fermions (type-III seesaw) instead of the $SU(2)_L$-singlet fermions (type-I). Being SM gauge multiplets, the type-II and type-III seesaw messengers can be directly produced via gauge boson couplings at colliders, and the non-observation of such processes sets a lower limit on these particles at hundreds of GeV scale. Similarly, extending the SM gauge group, e.g. to $U(1)_{B-L}$ or $SU(2)_L\times SU(2)_R\times U(1)_{B-L}$ for ultraviolet-completion of seesaw, opens up additional production channels for the seesaw messengers, and again, the lack of any positive signal sets stringent constraints on the seesaw scale in such cases~\cite{Deppisch:2015qwa}. 

An alternative explanation for small neutrino masses is that they arise only as quantum corrections. In these radiative neutrino mass models~\cite{Cai:2017jrq},
the tree-level Lagrangian does not generate the dim-5 LNV operator, owing to the particle content or
symmetries present in the model, but small
Majorana neutrino masses are induced only at the loop level. There exist both high-energy collider and low-energy charged lepton flavor violation constraints on the heavy particles going in these loops. In addition, the radiative models with at least one SM particle in the loop can lead to observable neutrino non-standard interactions~\cite{Babu:2019mfe}, thus making them testable in the near future.   
  
\vspace{-0.2cm}
\section{Connection to Other Puzzles}
\vspace{-0.2cm}
Neutrinos can also provide the missing link to other BSM phenomena. For example, the same sterile neutrinos responsible for neutrino mass could also explain the observed baryon asymmetry of the Universe via the mechanism of leptogenesis~\cite{Bodeker:2020ghk}. A large chunk of the allowed parameter space for leptogenesis can be tested at future colliders.  Similarly, the lightest sterile neutrino, if cosmologically stable, can be a good DM candidate~\cite{Drewes:2016upu}. In the left-right symmetric extension, the neutral component of the $SU(2)_R$-triplet scalar field can be an alternative decaying DM candidate~\cite{Dev:2025fcv}. The best signature of these DM candidates is a nearly monochromatic line feature in $X$-ray and gamma-ray observations.   

\vspace{-0.2cm}
\section{Conclusion}
\vspace{-0.2cm}
Neutrinos have given us the first (and so far only) laboratory evidence of BSM physics. Even in the precision era of neutrino oscillation physics, there are many open questions in neutrino theory. To address them, it is necessary to cast a broad net covering multiple experimental frontiers and exploring potential connections to other BSM puzzles, 
including baryon asymmetry and dark matter.

\medskip
\noindent
{\bf Acknowledgments:} The work of B.D. was partly supported by the U.S. Department of Energy under grant No. DE-SC0017987.
\vspace{-0.2cm}
\bibliographystyle{utphys}
\bibliography{ref_dev.bib}
\end{document}